\def\year{2015}
\documentclass[letterpaper]{article}
\usepackage{aaai}
\usepackage{times}
\usepackage{helvet}
\usepackage{courier}
\usepackage{comment}
\usepackage{graphicx}
\usepackage{url} 
\begin{document}
\title{Detection of Cyberbullying Incidents on the Instagram Social Network}

\author{Homa Hosseinmardi, Sabrina Arredondo Mattson, Rahat Ibn Rafiq, Richard Han, Qin Lv, Shivakant Mishra \\
Computer Science Department \\
University of Colorado Boulder \\
Boulder, Colorado \\
}
\maketitle
\begin{abstract}
\begin{quote}
Cyberbullying is a growing problem affecting more than half of all American teens. The main goal of this paper is to investigate fundamentally new approaches to understand and automatically detect incidents of cyberbullying over images in Instagram, a media-based mobile social network. To this end, 
we have collected a sample Instagram data set consisting of images and their associated comments, and designed a labeling study for cyberbullying as well as image content using human labelers at the crowd-sourced Crowdflower Web site.  An analysis of the labeled data is then presented, including a study of correlations between different features and cyberbullying as well as cyberaggression. Using the labeled data, we further design and evaluate the 
accuracy of a classifier to automatically detect incidents of cyberbullying.
\end{quote}
\end{abstract}

%\noindent Congratulations

%\begin{itemize}
%\item You must use the latest AAAI Press \LaTeX{} macro.
%\end{itemize}

% !TEX root = main.tex
\section{Introduction}
\label{sec:intro} 

As online social networks (OSNs) have grown in popularity, instances of cyberbullying in OSNs have become an increasing concern.  In fact more than half of American teens have been the victims of cyberbullying \cite{stop}. Although cyberbullying may not cause any physical damage initially, it has potentially devastating psychological effects like depression, low self-esteem, suicide ideation, and even suicide \cite{Hinduja,Menesini}. Incidents of cyberbullying with extreme consequences such as suicide are now routinely reported in the popular press. For example, Phoebe Prince, a 15-year-old high school girl, committed suicide after being cyberbullied by negative comments in the Facebook social network \cite{Pheobe}. Hannah Smith, a 14-year-old, hanged herself after negative comments were posted on her Ask.fm page, a popular social network among teenagers \cite{hannah}. Cyberbullying is such a serious problem that nine suicides have been linked with cyberbullying on the Ask.fm Web site alone \cite{askfmsuicides}. Although cyberbullying is not the direct cause of these suicides,  cyberbullying was viewed as a contributing factor in the death of these teenagers \cite{cyber1}. 

\begin{figure}[!ht]
\centering
\includegraphics[width=0.3\textwidth]{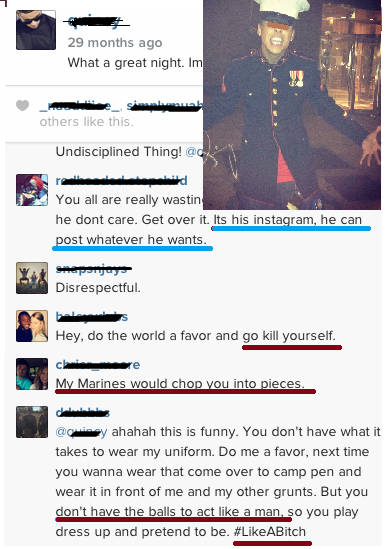}
\caption{An example of comments posted on Instagram.  To give more room for the text, we have moved the associated image to overlay some of the text.}
\label{fig:insta}
\end{figure}

Given the gravity of the consequences cyberbullying has on its victims and its rapid spread among middle and high school students, there is an immediate and pressing need for research to understand how cyberbullying occurs in OSNs today, so that effective techniques can be developed to accurately detect cyberbullying.
%prevent, mitigate and possibly even predict an onset of cyberbullying in online social networks. 
A recent survey on cyberbullying \cite{annual} has listed the top five networks where the highest percentage of users have reported experiencing cyberbullying, namely Facebook, Twitter, YouTube, Ask.fm, and Instagram.  Instagram is of particular interest as it is a media-based mobile social network, which allows  users to post and comment on images.  Cyberbullying in Instagram can happen in different ways, including posting a humiliating image of someone else by perhaps editing the image, posting mean or hateful comments, aggressive captions or hashtags, or creating fake profiles pretending to be someone else \cite{insta}. Figure \ref{fig:insta} illustrates an example of an attack in Instagram in which offensive hashtags and hateful comments were posted to humiliate the profile owner.

Cyberbullying has been defined as intentionally aggressive behavior that is \emph{repeatedly carried out} in an online context against a person \emph{who cannot easily defend him or herself} \cite{kowalski2012cyberbullying,patchin2012update}.  It is important to this definition of cyberbullying that both the frequency of negativity and the imbalance of power between the victim and perpetrator be taken into account.  In contrast, cyberaggression is a more general type of behavior that is broadly defined as using digital media to intentionally harm another person~\cite{kowalski2012cyberbullying}.  

Prior works that investigated cyberbullying \cite{Ptaszynski2010},\cite{Improved},\cite{usingML},\cite{Dinakar_modelingthe2},\cite{Twitter},\cite{kontostathis2013detecting}, \cite{xu2012learning},\cite{Nahar14},\cite{Nahar13},\cite{dinakar2011modeling},\cite{nahar2012sentiment} are more accurately described as research on cyberaggression, since these works do not take into account both the frequency of negativity and the imbalance of power.  These works have largely applied a text analysis approach to online comments, since this approach results in higher precision and lower false positives than simpler list-based matching of negative words \cite{sood2012profanity}.  Other work analyzed negativity in Ask.fm \cite{asonam14} and Instagram comments \cite{homa-cyberbullying-socialcom14}, but did not label the data at all.  %according to the cyberbullying definition above.   
%A key limitation of much of the prior work is that the proper definition of cyberbullying is not considered, i.e. neither the frequency of negativity nor the imbalance of power between the victim and bully is considered when labeling comments. It is more accurate to say that most of the prior work on cyberbullying is more appropriately described as research either on the topic of cyberaggression, which has a less restrictive definition as we discuss later in the labeling section, or on negative word analysis, wherein not all occurrences of negative words constitute cyberaggression, much less cyberbullying.

\iffalse For example, Phoebe Prince, a 15 year old high school girl, committed suicide after being cyberbullied by negative comments in the Facebook social network \cite{Pheobe}. Hannah Smith, a 14 year old, hanged herself after negative comments were posted on her Ask.fm page, a popular social network among teenagers \cite{hannah}. Cyberbullying is such a serious problem that nine suicides have been linked with cyberbullying on the Ask.fm Web site alone \cite{askfmsuicides}.   Although cyberbullying is not the direct cause of these suicides,  Cyberbullying was viewed as a contributing factor in the death of these teenagers \cite{ref1}. \fi 

\iffalse
 While most current studies have focused on the prevalence and impact of cyberbullying in education and psychology \cite{4,5,6}, our interest is in understanding how social networks are being used to enable cyberbullying. This paper will identify and characterize cyberaggression (intentional negative electronic behavior) and will explore the extent to which we can identify and characterize cyberbullying via online posted media.  \fi
 
Additional research investigated aspects of the Instagram social network, but not in the context of cyberbullying.  For example, \cite{Weilenmann} explored users' photo sharing experience in a museum.  \cite{insta} considered the temporal photo sharing behavior of Instagram users.  \cite{whatinsta} categorized Instagram images into eight popular image categories and the Instagram users into five types in terms of their posted images. By investigating user practices in Instagram, \cite{araujo2014} concluded that users tend to be more active during weekends and at the end of the day. They also found out that users are more likely to like and comment on the medias that are already popular, thereby inducing the rich get richer phenomenon.
%We aim at finding when a media has been shared in instagram, whether it has been targeted for cyberagression/cyberbullying.
 
In this paper, we make the following contributions:
 
\begin{itemize}
	\item We provide a clear distinction between cyberbullying and general
cyberaggression. Cyberbullying is one type of cyberaggression, 
and most of the earlier research in this area has focused on identifying
cyberagression, which is relatively easier than identifying cyberbullying.
	\item We investigate cyberbullying behavior in Instagram by labeling collected Instagram images and their associated discussion comments according to both the more restrictive definition of cyberbullying and the more general definition of cyberaggression.  
	\item We present an analysis of the labeled images and comments, including the relationships of cyberbullying and cyberaggression to a variety of features, such as number of associated comments, N-grams, followed-by and following behavior of the posting users, liking behavior, frequency of comments, and labeled image content.
	\item We design and evaluate multi-modal classifiers to detect cyberbullying based on the labeled data, measuring accuracy across different feature sets including text, images, and meta data.
%	Definition, data collection and labeling: Develop an understanding of what is cyberbullying in the context of Instagram, a popular mobile-based image sharing social network, collect relevant multimodal texts and images, and label this data into categories such as cyberaggression and cyberbullying incident 
%	\item Analyze and characterize Instagram user behaviors in the context of cyberaggression and cyberbullying.
%	\item Develop multimodal fusion classifiers to automatically and accurately detect incidents of cyberbullying.
\end{itemize}

%{\bf INCLUDE A Paragraph about key findings of this paper here}

% !TEX root = main.tex
\section{Data Collection}
Using a snowball sampling method, we have identified 41K Instagram user ids.  
%We collected data for 41K Instagram users using a snowball sampling method.
%We collected a comprehensive set of user data from the Instagram social network. For Instagram users, starting from a seed node, 41K user ids were gathered with a snowball sampling method. 
%Among these Instagram ids, there were both public and private profiles. 
61\% of these Instagram ids have public profiles, which is about 25K
public profiles. These 25K public user profiles comprise our complete set of typical Instagram users data. For each public Instagram user, the collected profile data includes the media objects/images that the user has posted and their last 150
associated comments, user id of each user followed by this user, user id of each user who follows this user, and user id of each user who commented on or liked  the
media objects shared by the user. We consider each media object/image and 
its associated comments as a {\em media session}. 
%To gain deeper insights, we divided all user posts into two groups, posts by profile owners and posts by other non-owner users, whom we will loosely term "friends" in the remainder of this paper though some of these other users may not be truly friendly. As the goal is finding instances of cyberbullying, 
For this set of 25K users, 697K media sessions were collected.

In order to make the labeling of cyberbullying more manageable, we sought to label a smaller subset of these media sessions. 
% that included the media and the comments associated with it.  
We focused on those media sessions that have a high percentage of negativity in their associated comments, since we reasoned that this should give us a higher likelihood of identifying cyberbullying once the data was properly labeled. We used the same approach as in the previous works \cite{asonam14,homa-cyberbullying-socialcom14} for tagging a comments as a negative or not, by looking for profanity words coming from a dictionary obtain form  \cite{noswearing,NegativeWordsList}.
%Labeling cyberbullying is costly and time consuming process. For As we are not able to label all the images,  we look for a small set of the media objects which have the high percentage of negativity in their posts to have higher chance of catching cyberbullying when labeling the data. 
Specifically, we select images using the following two criteria: 
%The selected images have the following properties:

\begin{itemize}
  \item the media has at least 15 comments, and
  \item more than 40\% of the comments by users other than the profile owner have at least one negative word.
\end{itemize}

Using these criteria, we were able to reduce the number of media sessions to a more tractable group of about $1,203$. When we returned to  Instagram to collect images associated to the comments of selected media's for labeling, only 998 were still available, for the rest either the media session was deleted or the profile was made private or deleted.  The reason for putting lower bound on the number of comments (minimum 15 comments) is to ensure that there are enough comments to adequately assess
the frequency or repetition of negativity, which is an important part of the 
cyberbullying definition. The average number of comments per image is 59.6,
and we decided to analyze all images with number of comments at least a
quarter of this average number. For these $998$ media sessions, the average number of comments associated with a media is about $64.3$.

%Each image has on average 59. 6 comments. We kept images their number of comments is bigger than 0.25 of the average number of comments. The reason for considering a lower bound is to have enough comments to catch the repetition part of cyberbullying definition and avoid cases with only few comments. %We collected with this 1100 image ids, however when we came back to collect images belongs to the set media ids, some where deleted and we could only collect 998 images. 

\begin{figure}[!ht]
\centering
\includegraphics[width=0.5\textwidth]{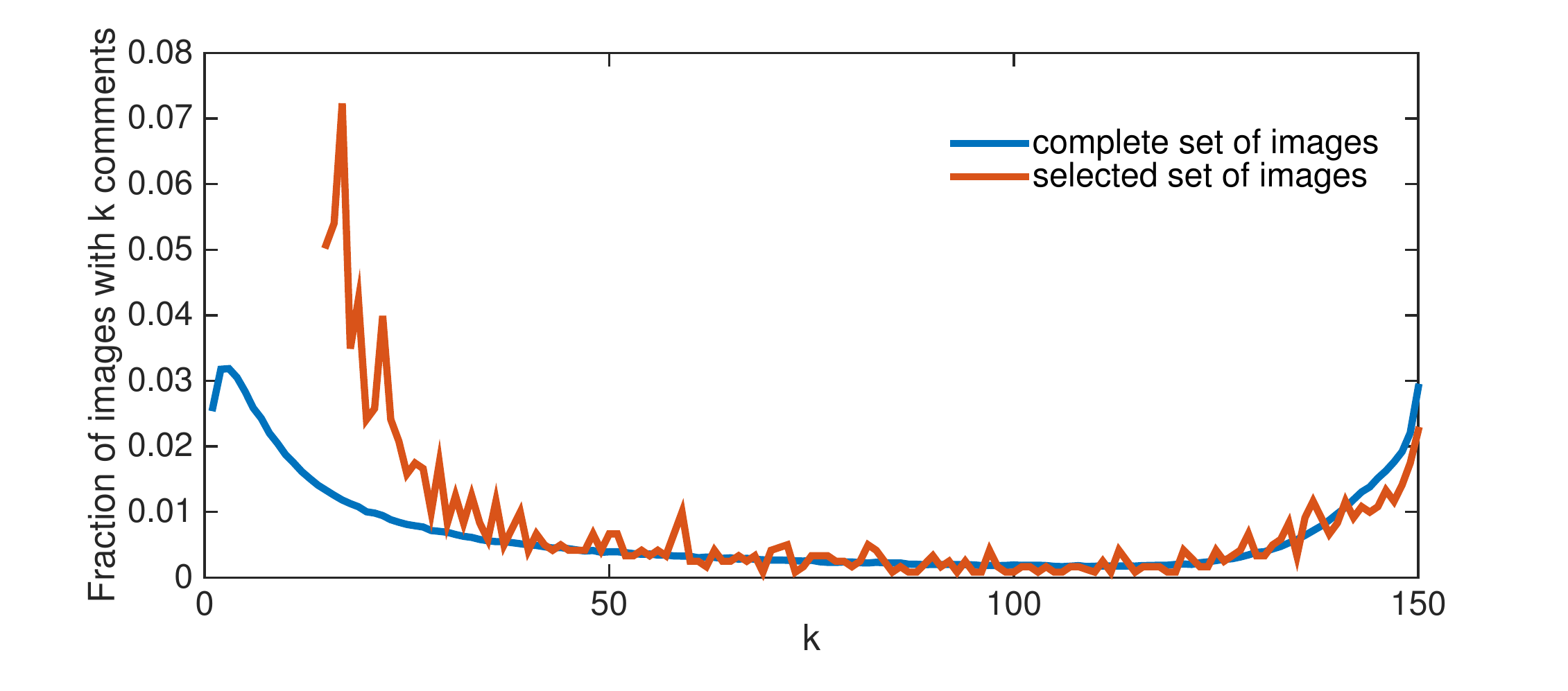}
\caption{Comparison of the distribution of the number of comments per collected Instagram media session. Blue is for the complete set of media sessions, and red is for the selected subset of 998 media sessions with more than 15 comments and high degree of negativity.}
\label{fig:comment_dist}
\end{figure}

Figure \ref{fig:comment_dist} shows the distribution of the number of comments for our selected smaller subset of media sessions compared with the number of comments for the complete set of media sessions. We observe that the fraction of images with number of comments between 15 and 50 is higher in the selected data set than that in the complete set. However, the distribution is similar when the number of comments is greater than 50. This shows that media sessions with relatively higher negativity tend to be confined to moderate number of comments.

\begin{figure}[!ht]
\centering
\includegraphics[width=0.5\textwidth]{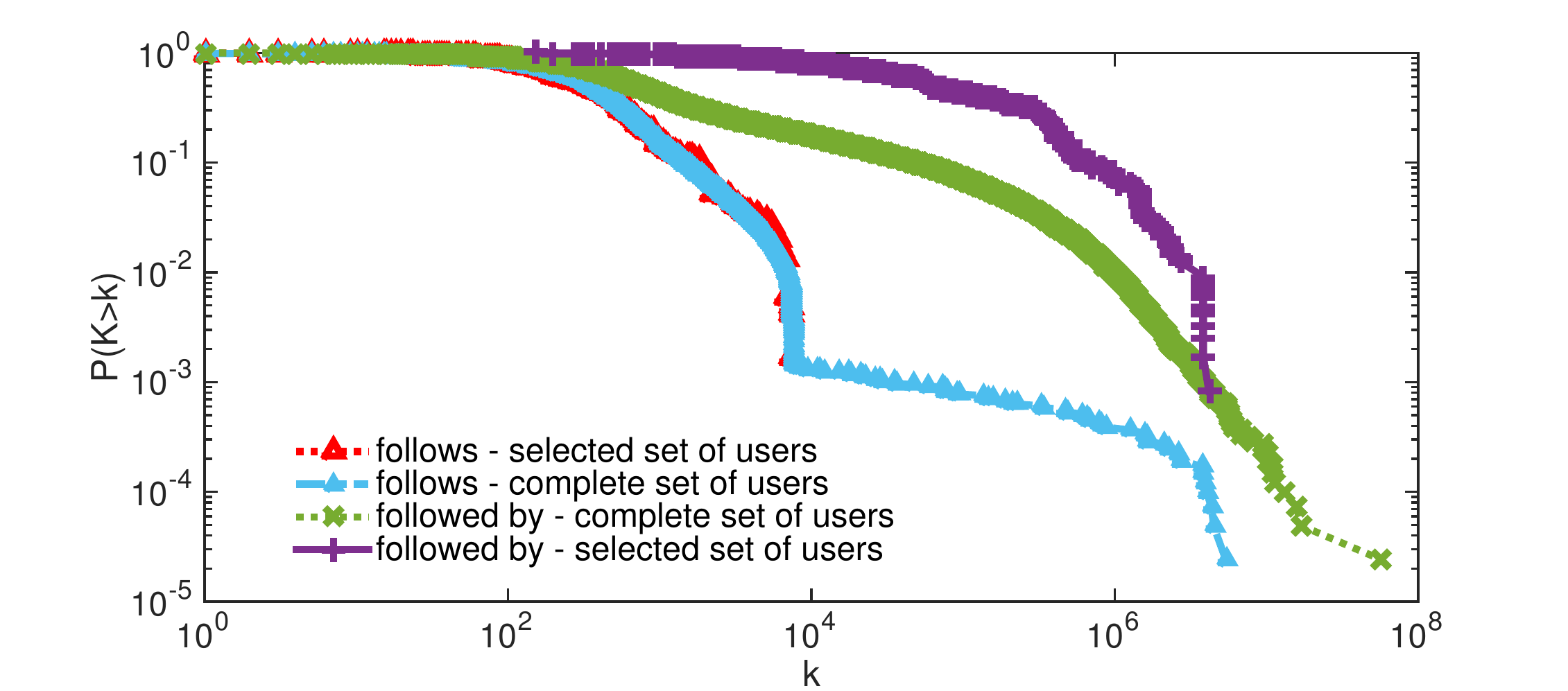}
\caption{CCDF of the number of followed by and follows for users in the complete set and highly negative subset of media sessions.}
\label{fig:repeated}
\end{figure}

Figure~\ref{fig:repeated} illustrates the CCDF of the number of followed by and follows for users in both the complete and selected set of media sessions.  We see that the
number of follows for users in the complete and selected sets exhibit the same pattern. However, the distribution for selected users ends at around 7,500, while the distribution for all users goes to almost $10^7$. 
On the other hand, distributions of the number of followed by users are
different for selected users and all users.
The number of followed by users is higher for the selected users, but this
distribution ends at around $4.16 * 10^6$, while the distribution for all users
goes all the way up to $10^8$. Looking at the data more closely, we found that
a large number of images posted by the selected users set correspond to popular
personalities or events, e.g., a lot of these users are singers, celebrities,
tattoo artists, or simply users who are popular within a local area. These
users draw a lot of attention. Because of their popularity,
they have a relatively larger number of followers, and tend to 
attract a significant number of negative comments in the form of criticism
from other users.

%{\bf IN All figures, change `normal' to `all'}

% !TEX root = main.tex
\section{Cyberbullying Labeling}

In this section, we explain the design and methodology for our survey labeling the selected set of media sessions. Our first challenge is choosing appropriate definitions of terms, which will then be used in ground truth labeling.  Based on the literature, a major early choice that we have made is to distinguish between cyberaggression and cyberbullying.  Cyberaggression is broadly defined as using digital media to intentionally harm another person~\cite{kowalski2012cyberbullying} \iffalse "intentionally posting harmful or negative content" in online forums such as social networks"~\cite{cyberaggdef} \fi.  Examples include negative content and words such as profanity, slang and abbreviations that would be used in negative posts such as hate, fight, wtf.
Cyberbullying is one form of cyberaggression that is more restrictively defined as intentional aggression that is repeatedly carried out in an electronic context 
%such as e-mail, blogs, instant messages, text messages, and posts on social networks 
against a person who cannot easily defend him or herself \cite{kowalski2012cyberbullying,patchin2012update}. 
%Cyberaggression is more broadly defined as using the internet or other technologies to intentionally harm another person \cite{kowalski2012cyberbullying}.
Thus, cyberbullying  \iffalse and traditional bullying share \fi consists of three main features \iffalse that are important when conceptualizing and measuring cyberbullying \fi: (1) an act of aggression online; (2) an imbalance of power between the individuals involved; and (3) it is repeated over time \cite{hunter2007perceptions,kowalski2012cyberbullying,olweus1993bullying,olweus2013school,Smith2012}. The power imbalance can take on a variety of forms including physical, social, relational or psychological \cite{dooley2009cyberbullying,monks2006definitions,olweus2013school,pyzalski2010electronic}, such as a user being more technologically savvy than another~\cite{kowalski2014bullying}, a group of users targeting one user, or
a popular user targeting a less popular one \cite{Limber2008}. Repetition of cyberbullying can occur over time or by forwarding/sharing a negative comment or photo with multiple individuals \cite{Limber2008}.

\begin{figure}[!ht]
\centering
\includegraphics[width=0.5\textwidth]{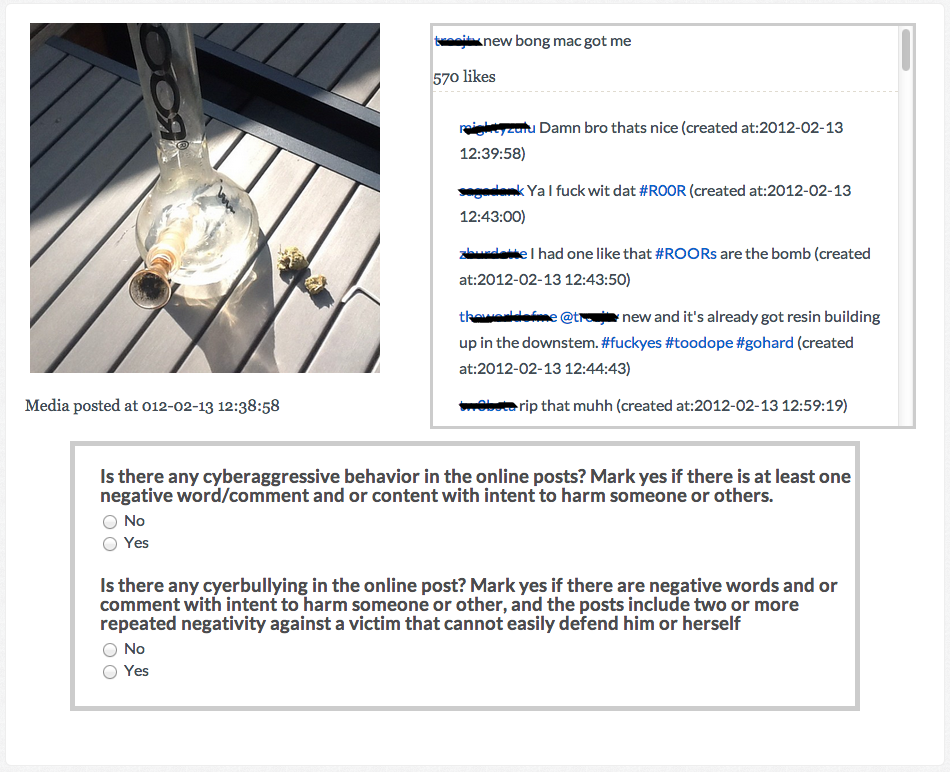}
\vspace{-2.0em} 
\caption{An example of the labeling survey, which shows an image and its corresponding comments, and the survey questions.}
\label{fig:survey}
\end{figure}

\begin{figure*}[lht]
\centering
\includegraphics[width=0.8\textwidth]{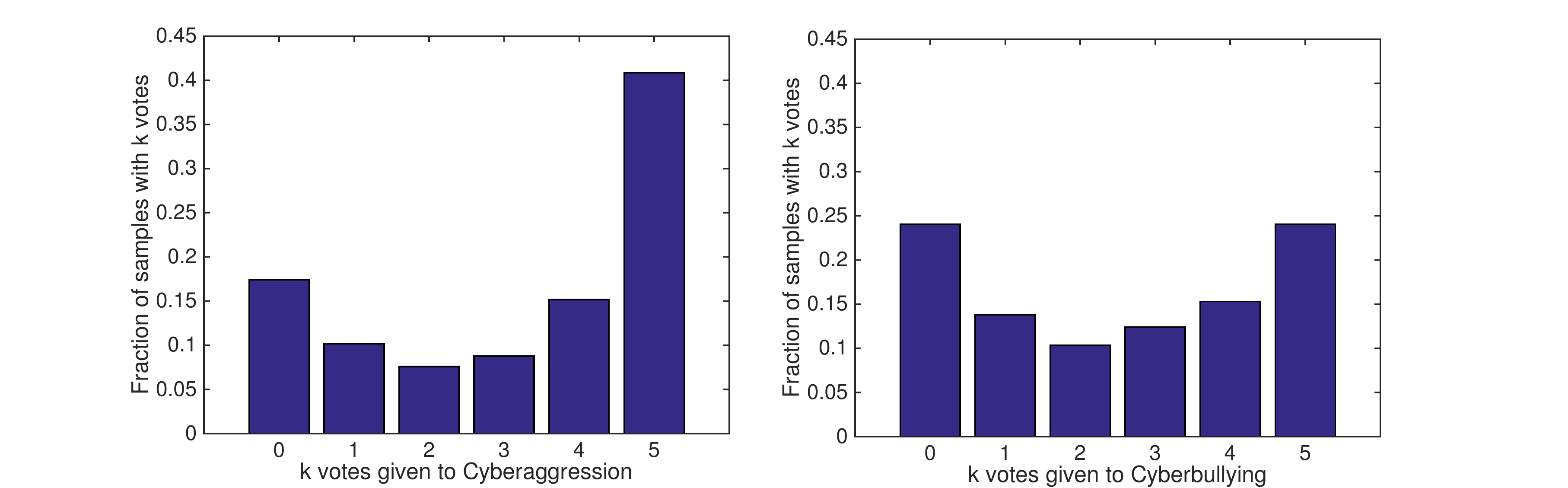}
\vspace{-1.0em}
\caption{Fraction of media sessions that have been voted $k$ times as cyberagression (left) or cyberbullying (right).}
\label{fig:dist_votes}
\end{figure*}

In Instagram, each media session consists of a media posted by the profile owner and the corresponding comments for the media object. The goal in this paper is to investigate cyberaggression and cyberbullying in this multi-modal (textual comments and media objects) context.  Therefore, the design of our survey needed to incorporate both the image and the associated text comments when asking the human labeler whether the media session was an instance of cyberbullying or cyberaggression.  Figure~\ref{fig:survey} illustrates an example of our design for the labeling survey.  On the left is the image, and on the right is a scrollable interface to help the labeler see all of the comments associated with this image.  
With the help of an expert, we decided to ask the labelers two questions, namely whether the media session constituted cyberaggression or not, and whether the media session constituted cyberbullying or not.  During the instructional phase prior to labeling, labelers were given the aforementioned definitions of cyberaggression and cyberbullying along with related examples.  Each media session was labeled by five contributors.
%In the designing the survey, after explaining above definitions briefly,  we ask contributors to answer the question whether cyberagressive or cyberbullying interaction has been received by the post Figure~\ref{fig:survey}. They could say no if non of the later answers apply to the post. Each post was labeled by five contributors.

%With the help of expert, and trying several surveys with small datasets, we finalized our survey with asking two questions.  First whether the contributor observe any aggressive interaction, second whether they observe repetition and imbalance of power in the aggressive behavior which can be called cyberbullying. Instruction with the complete definitions and examples were provided. 

To monitor the quality of labeling, potential contributors were given the answers to a set of examples, and then were subjected to a pre-filtering step in which they were asked to answer a set of similar quiz questions.  Contributors needed to answer correctly a minimum number of quiz questions to qualify as a labeler for our survey.  Also during the job, random test questions were asked to monitor the quality of the labeling during the job. A minimum threshold amount of time was also set to filter out contributors who rushed through the labeling without spending a sufficient minimum duration to ensure the quality of the labeling.

We were also interested in image contents of media sessions that had been targeted with a high proportion of negative comments.  If the type or category of an image could be identified, then this may prove to be a useful feature in classification of cyberbullying.  We first sampled some of the images in the selected subset to determine a suitable set of representative categories or types to be used in the labeling.  For example, some of the dominant categories were the presence of a human in the image, as well as animals, text, clothes, tattoos, sports and celebrities. We then conducted a second survey focused only on the image content, and asked labelers to identify which of the aforementioned categories were present in the image.  Multiple categories could be selected for a given image.
%designed our survey based on these  limited set of categories and provided a list of multiple choices for the content images.

\iffalse
\begin{table*}[htbp!]
\centering
\begin{tabular}{|c|c|c|c|c|c|c|c|c|c|c|}
\hline 
content  &  	 car & cartoon & cloths & drugs & food & human & animal & tattoo & text & other	\\ \hline
number	  & 22 & 14 & 33 & 19 & 11 & 500 & 6 & 40 & 286 & 94 \\ \hline
\end{tabular}
\caption{Image content}
\label{stats}
\end{table*}

\fi

% !TEX root = main.tex

\section{Analysis and Characterization of Labeled Cyberbullying}

\begin{table*}[htbp!]
\centering
\begin{tabular}{ |l | c | c|c | c | c | c| }
\hline 
    \hline
    correlation & likes & media & followed by & following  \\ \hline
   cyberbullying  & 0.069  & 0.04 & 0.17& -0.02\\ \hline
    cyberaggression & 0.04 &0.07 & 0.14& -0.00\\
    \hline
\end{tabular}
\caption{Correlation between number of votes for cyberbullying /cyberagression and image/user meta data}
\label{stats1}
\end{table*}

\iffalse
\begin{table*}[htbp!]
\centering
\begin{tabular}{ |l | c | c | c | c | c| }
\hline 
    \hline
    correlation & likes$^*$ & media & following$^{**}$ & followed by$^{***}$ \\ \hline
   non cyberbullying  & 4262.12 &  1176.33 & 778.01 & 188034.94  \\ \hline
    cyberbullying  & 5753.37 &  1187.180 & 666.97& 382378.92  \\ \hline
    \hline
\end{tabular}
\caption{Mean values of image/user meta data for cyberbullying versus not. ($^* p<0.05$,$^{**} p<0.1$, $^{***} p<0.0001$) }
\label{stats1}
\end{table*}

\begin{table*}[htbp!]
\centering
\begin{tabular}{ |l | c | c | c | c | c| }
\hline 
    \hline
    correlation & likes & media &  following  & followed by$^{*}$\\ \hline
   non cyberagression  & 4481.08 &  1140.51 & 789.28 & 179676.42 \\ \hline
    cyberagression  & 5236.64 &  1218.81 & 692.94 & 347100.67 \\ \hline
    \hline
\end{tabular}
\caption{Mean values of image/user meta data for cyberaggression versus not. ($^* p<0.0001$)}
\label{stats2}
\end{table*}

\fi

We submitted our survey with 998 media sessions (images and their associated comments) to CrowdFlower, a crowd sourcing website, each labeled by five different contributors.  Figure~\ref{fig:dist_votes} illustrates the distribution of the labeled answers among the five labelers for each of the two questions on cyberbullying and cyberaggression.  The higher the number of votes for a given label, the more confidence that we have that a given media session constitutes cyberaggression or cyberbullying, where five votes constitutes unanimous agreement.
\iffalse
In the process of labeling, high agreements shows better quality. However judgment of a sample for aggression or bullying is a subjective problem which depends on many factors. So, we consider every image-comments that has been labeled as
aggression/bullying, even if by just one contributor, as this labeling
indicates that there is some probability that the image-comments is harmful. 
\fi
The left chart in Figure  \ref{fig:dist_votes} shows the percentage of samples that have been labeled as cyberaggression $k$ times, and the right chart shows the percentage of samples that have been labeled as cyberbullying $k$ times.  

We notice that for cyberaggression, most of the probability mass
is around media sessions labeled as cyberaggression by all five contributors.
This is not surprising since all the samples have at least 40\% negative comments. However, we observe that around 17\% of the media sessions have not been labeled as cyberaggression at all by any of the five contributors. This suggests that only employing a high percentage of negativity threshold of 40\% to detect cyberaggression can still produce many false alarms.

For cyberbullying labeling (right chart in Figure~\ref{fig:dist_votes}), we notice that about 24\% of the
media sessions have not been labeled as cyberbullying by any of the
five contributors, even though these samples were originally selected for their high negativity.
Further, we observe that about 48\% of the media sessions have two or fewer votes.  If we apply a majority vote criterion to deciding whether a given session was cyberbullying or not among the five labelers, then nearly half of the sessions would be defined as not cyberbullying, despite their high percentage negativity.
\emph{Therefore, a key finding of our labeling is that a large fraction of Instagram media sessions with a high percentage of negative words would \textbf{not} be deemed as cyberbullying.} 
\iffalse
 while only about 48\% of the samples have been labeled as
cyberbullying by four or all five contributors.  The lowest percentages
belong to image+comments samples that have been labeled as cyberbullying
by two or three contributors.
\fi
The implication is that classifier design for cyberbullying here cannot solely rely on the percentage of negativity among the words in the image-based discussion, since this would produce many false positives, but instead must consider other features to improve accuracy.

\emph{Another key observation is that the labelers are mostly in agreement about what behavior constitutes cyberbullying and what does not in Instagram media sessions.}  That is, most of the labelers agree either that a media session is cyberbullying (about 40\% of sessions have either four or all five votes for cyberbullying) or that it is not (about 38\% of sessions have either zero or one vote).  Only about 22\% of media sessions have two to three votes, so there is some disagreement in a small fraction of cases about whether the session is cyberbullying or not.

\begin{figure}[!ht]
\centering
\includegraphics[width=0.3\textwidth]{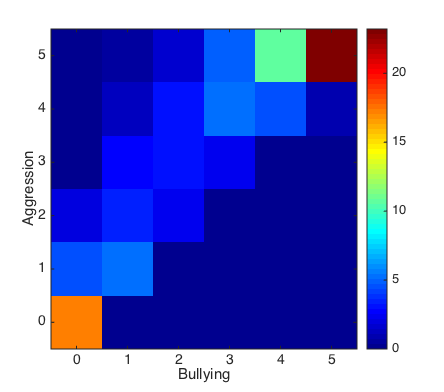}
\caption{Two-dimensional distribution of media sessions as a function of the number of votes given for cyberagression versus the number of votes given for cyberbullying, assuming five labelers.}
\label{fig:heat1}
\end{figure}

In order to understand the relationship between labeled cyberaggression and labeled cyberbullying media sessions, we plotted in Figure~\ref{fig:heat1} a two-dimensional heat map that shows the distribution of media sessions as a function of the number of votes each media session received for cyberaggression and cyberbullying.  We observe that a significant fraction of the sessions exhibit strong agreement in terms of both receiving high numbers of votes for both cyberbullying and cyberaggressions, or both receiving low numbers of votes, i.e. the session is neither cyberbullying nor cyberaggression.  This is shown by the high energy in the upper right and lower left along the diagonal.  In addition, it is promising that the area below the diagonal is essentially zero, meaning no sessions have received more votes for cyberbullying than cyberaggression.  This conforms with the definition that cyberbullying is a subset of cyberaggression.  

We see that the remaining significant energy in the distribution appears in the area above the diagonal.  Media sessions in this area exhibit the property that if they receive $N_1$ cyberbullying labels, then they receive $N_2 \geq N_1$  cyberaggression labels.  This area corresponds to cases where there is cyberaggression but not cyberbullying.  In particular, if we look at the cases where there is some disagreement as to whether a session is cyberbullying or not ($N_1=2$ or 3 votes for cyberbullying), we see that there is significant support that these sessions exhibit cyberaggression (there is significant energy for $N_2$ values of four and five votes for cyberaggression).  In fact, the dominant value for cyberagression when $N_1=2$ is $N_2=4$, and similarly the dominant value for cyberaggression for $N_1=3$ is $N2=4$ or 5.  \emph{As a result, our analysis is able to quantify that there is substantial support for identifying Instagram media sessions that exhibit cyberaggression but not cyberbullying.}

%However observing non zero values in the upper diagonal shows example of negativity which have more cyberaggression but less for being cyberbullying. 

\iffalse
In fact, since cyberbullying is a subset of cyberaggression, we expect that 
if an image has not been labeled as cyberaggression at all, then it
should not be labeled as cyberbullying.
{\bf Need some statistics about images labeled as cyberaggression but not
cyberbullying here.}

In order to validate the quality of the labeling, we provide heat-map Figure 6 which shows what percentage of images have been voted k times as cyberagression and j times as cyberbullying. We can see that the majority of the samples has been voted as strong agreement on not-cyberbullying/not-cyberaggression or cyberbullying/cyberaggression. It is promising that the lower diagonal is zeros, meaning no samples has received more votes for cyberbullying than cyberaggression. However observing non zero values in the upper diagonal shows example of negativity which have more cyberaggression but less for being cyberbullying. 
\fi

Next, we would like to examine the correlation between the strength of labeled cyberbullying/cyberaggression and a variety of other factors.  We define the strength of cyberbullying as the number of votes received for labeling a media session as cyberbullying, and similarly for cyberaggression.  Table~\ref{stats1} shows the correlation between the strength of cyberbullying/cyberaggression and media properties such as the number of likes, as well as meta data about the profile owner of the shared media object, such as the number of followings, followed-by's, and total shared media.  We observe that there is a correlation of 0.17 with the number of followed-by's while there is no significant correlation with the number of likes, total shared media, and followings.  We also found that cyberaggression exhibits similar but slightly weaker correlations to the same factors.

\begin{figure}[!ht]
\centering
\includegraphics[width=0.45\textwidth]{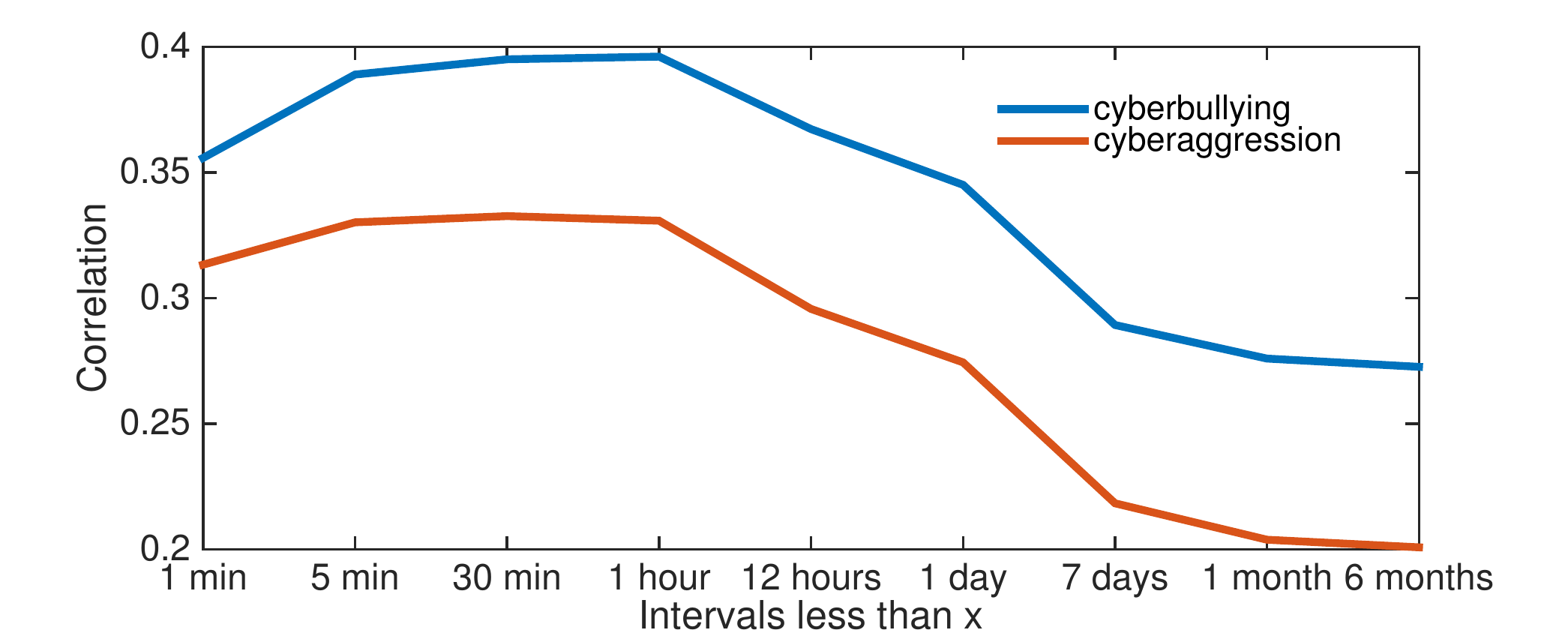}
\caption{Temporal correlation analysis.}
\label{fig:temporal}
\end{figure}

Our analysis further explores Pearson's correlation by considering temporal factors.  We would like to understand how the human labelers incorporated the definition of cyberbullying, which includes the temporal notion of repetition of negativity over time, into their labeling.  Given time stamps on every collected comment, we compute the interval or interarrival time between a comment and the next comment.  We then counted the number of comment interarrival times in a media session less than some threshold value.  Figure~\ref{fig:temporal} describes our results, namely that there is a strong correlation of about 0.4 between the strength of support for cyberbullying and media sessions in which there are frequent postings within 1 hour of each other.  Further, we find that as we expand the allowable duration between comments, that is comments are allowed to be further apart in time, then the correlation weakens considerably between more widely separated comments and support for labeling this session as cyberbullying.  We also considered cyberbullying's correlation with other temporal factors such as the median, mean and variance, i.e. jitter, of the comment interarrival times but found little correlation.  Cyberaggression temporal correlations follow a similar pattern.

\iffalse
\begin{table*}[htbp!]
\centering
\begin{tabular}{ |l | c | c|c | c | c | c|c | c|c | }
\hline 
    \hline
    correlation & num delta < 5 min  &  num delta < 1 hour & num delta < 1 day & median & mean & variance & minimum &  maximum  \\ \hline
   cyberbullying  & 0.39                 & 0.39                      &        0.34            & -0.08    & -0.16   &   -0.07 &  -0.19   &    -0.17   & -0.01  \\ \hline
   cyber-aggression & 0.33            & 0.33&0.27            & 0.14                   & -0.10   &   -0.12 &  -0.03   &    -0.17   & -0.00     \\
    \hline
\end{tabular}
\caption{Correlation between temporal commenting behavior and number of votes given to cyberbullying and cyberaggresion}
\label{stats2}
\end{table*}
\fi

\begin{figure*}[ht]
\centering
\includegraphics[width=1\textwidth]{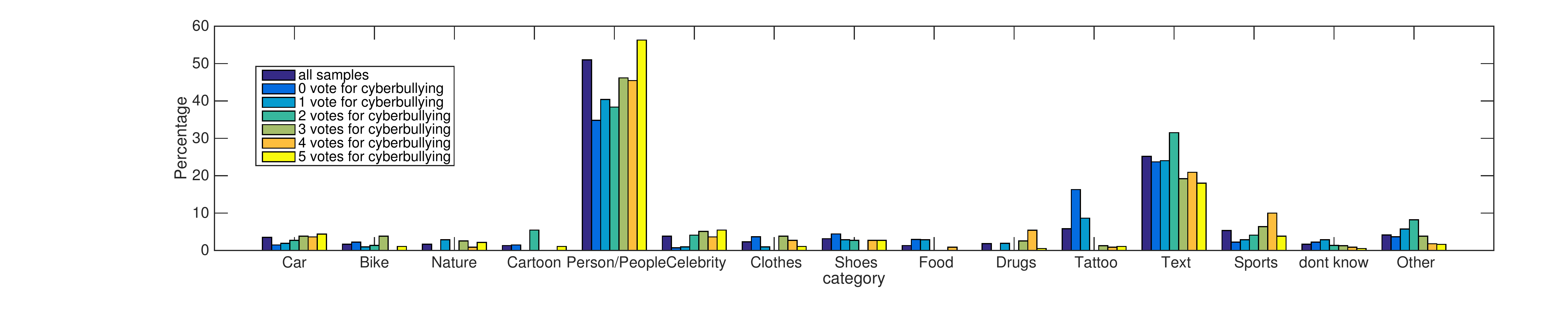}
\caption{Distribution of image categories given the media sessions have been voted for k times for cyberbullying. As some images belong to more than one category, the bars will sum up to more than one. }
\label{fig:bll}
\end{figure*}

\begin{figure*}[!ht]
\centering
\includegraphics[width=1\textwidth]{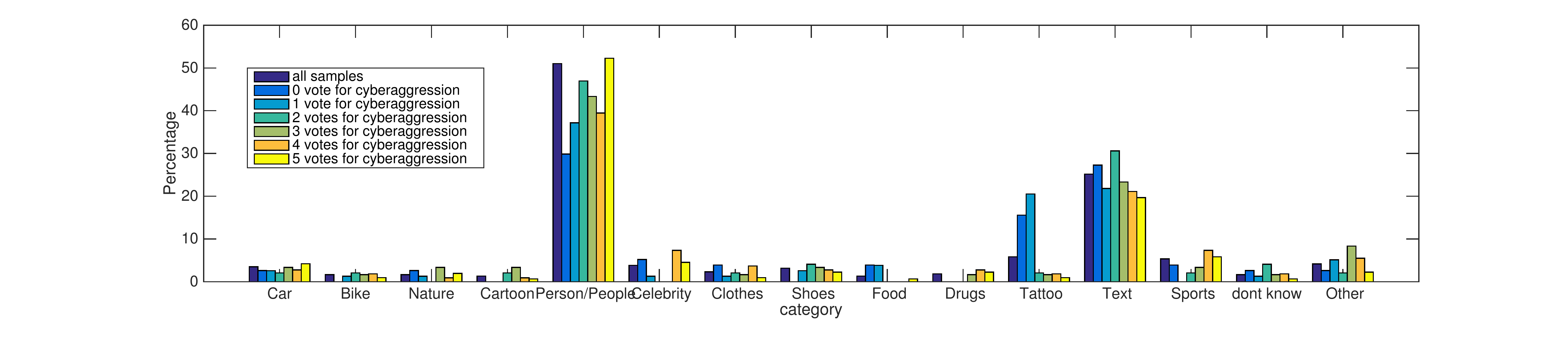}
\caption{Distribution of image categories given the media sessions have been voted for k times for cyberaggression.}
\label{fig:agg}
\end{figure*}

\iffalse
In order to find the correlation between temporal behavior of posting comments with cyberbullying/cyberaggression, in figure \ref{fig:temporal} we have calculated the correlation between number of votes for cyberbuulying and cyberaggression for number of comments which have been posted within less tan k minutes after their previous comment. The highest correlation is 0.397 for k equal to 60 minutes. 
\fi

\iffalse
If we consider k votes for cyberagression/cyberbullying means as how strong is it, we are interested to is the any correlation between media properties, such as number of likes and comment which the image receives, or some meta data about the profile owner of the shared media, such as followings, followed by and total shared media. Table \ref{stats} shows likes, followed by and following have no correlation with how strong an image-comment has been voted for cyberbullying or cyberagression. Higher correlations can be seen for number of comments for the images and number of followed by.  
\fi

\emph{To summarize, we have found that there are strong correlations between the strength of support for labeled cyberbullying and the number of text comments as well as the temporal property of the number of comments that are posted within one hour of one another in an Instagram media session.}

% !TEX root = main.tex

\subsection{Image Labeling Analysis}

In this subsection, we would like to understand the relationship between image content and cyberbullying in a media session.  Towards this end, we display the distribution results of our second survey on labeling image content in Figures~\ref{fig:bll} and ~\ref{fig:agg}.  First, we observe that among the media sessions with the highest negativity, the most common labels for the image content in these media sessions are Person/People, Text, Sports, and perhaps Tattoo, for most values of support for cyberbullying.  Second, there is some skew in distributions for certain labels such as Person/People, Tattoo and Sports, as the amount of support for cyberbullying varies.  For example, for images labeled as containing a Tattoo, we see a strong skew towards lower values of cyberbullying.  Such a skew may be helpful in classifier design, since whenever a tattoo is present, there appears to be little support that there is cyberbullying occurring, while whenever there is strong support for cyberbullying, images with tattoos are more scarce.  For Person/People, we see a skew in the opposite direction towards more cyberbullying support, and similarly for Sports.  Similar behavior is exhibited for cyberaggression as well.

%Normalizing number of images marked for each category to the total number of images, will cause the fractions sum up to more than one. 

\iffalse
We are also interested to see what is the fraction of images from each category, given the fact that image gets k vote for cyberbullyng \ref{fig:bll} or cyberagression \ref{fig:agg}. For some categories like tattoo we can see the distribution is skewed towards lower votes for cyberbullying/cyberagressin and for human category we can see higher fractions for larger values of votes for cyberbullying. Categories like Nature or clothes which have a symmetric distributions for 0 to 5 votes. Sport has more skewed distribution for cyberbullying compare to cyberaggression.  

From Figure \ref{fig:img}, we can see for some categories like tattoo a high percentage of images belong to None-Aggression/Bullying group, while the opposite is correct for celebrity, drugs and car. 
\fi

\begin{figure}[!ht]
\centering
\vspace{-1.0em} 
\includegraphics[width=0.35\textwidth]{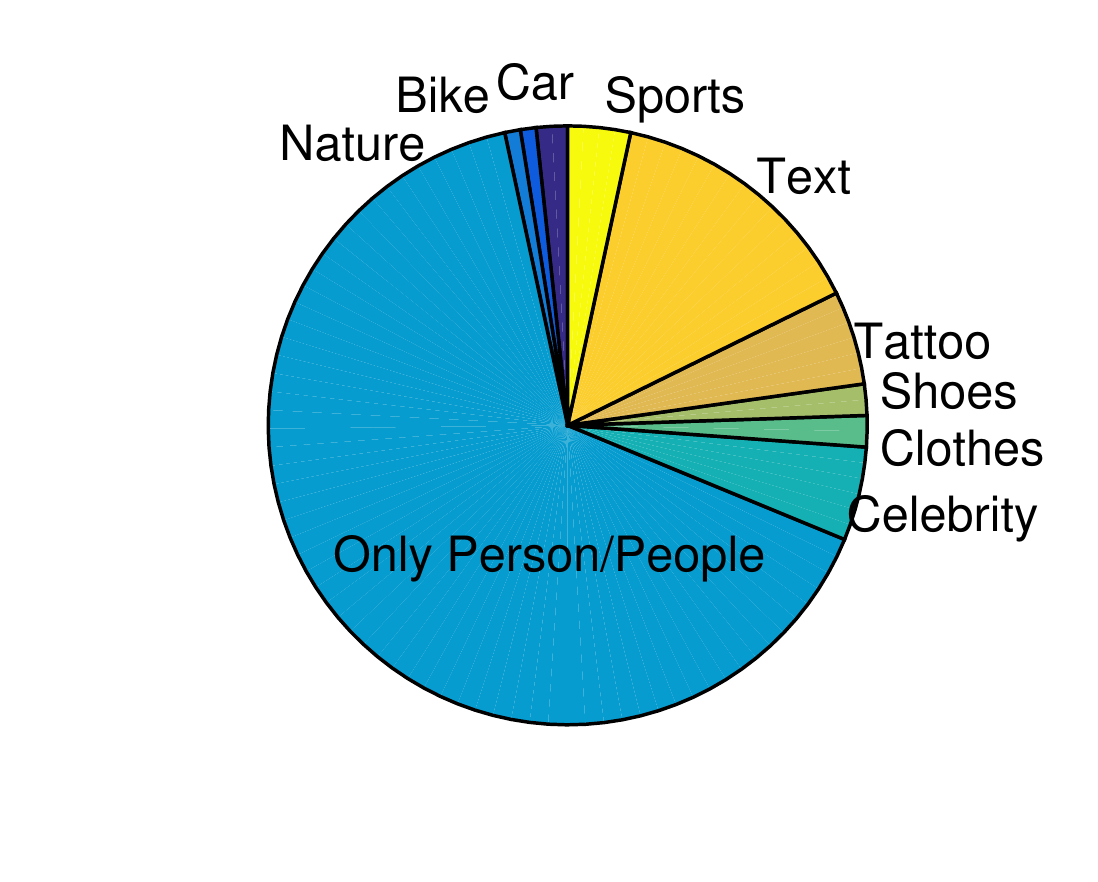}
\vspace{-3.0em} 
\caption{Fraction of other categories assigned to an image given that the image has been labeled as Person/People. }
\label{fig:imghuman}
\end{figure}

Since labeling of image content into more than one category was permitted, then we are further interested to see the distribution of multi-label images.  Figure~\ref{fig:imghuman} \iffalse and ~\ref{fig:imgtext} \fi shows the fraction of other categories assigned to a Person/People labeled image. \iffalse and a Text-labled image respectively \fi  For example, Figure~\ref{fig:imghuman} shows that more than 60\% of images labeled with Person/People were exclusively labeled as such, but about 15\% of such images were also labeled with the Text label.  Very few images were labeled with three labels. % Figure~\ref{fig:imgtext} shows that Text-Person/People labeling dominates any joint labeling with Text.

\iffalse
As some images belong to more than one category, the bars will sum up to more than one. For two categories text and person/people, we will look at what are the other categories which has been seen. 
\fi

\iffalse
\begin{figure}[!ht]
\centering
\includegraphics[width=0.5\textwidth]{./figure/pie_text-eps-converted-to.pdf}
\caption{Fraction of other categories assigned to an image given that the image has been labeled as Text.}
\label{fig:imgtext}
\end{figure}

\begin{figure}[!ht]
\centering
\includegraphics[width=0.2\textwidth]{./figure/celeb.jpg}
\caption{An example of an image labeled as celebrity, drugs and human.}
\label{fig:celeb}
\end{figure}
\fi

% !TEX root = main.tex
\section{Classifier Design and Evaluation}

To design and train the classifier, we chose to apply a majority vote criterion on the labeled data to determine whether a media session was cyberbullying or not.  Further, CrowdFlower provides us with a degree of trust for each labeler based on the percentage of correctly answered quiz and test questions during the labeling session.  This trust value is incorporated by CrowdFlower into a weighted version of the majority voting method called a ``confidence level".  We decided to employ this weighted trust-based majority voting metric as the basis for our classifier design.  Media sessions whose weighted trust-based metric was equal to or greater than 60\% were deemed to be strong enough support for cyberbullying.  Actually, 90\% of the original pure majority-vote based media sessions wound up in this higher-confidence cyberbullying-labeled group.  For this higher-confidence data set, 52\% in total belonged to the ``bullying" group while 48\% were not deemed to be bullying.  This provides a base case from which to compare our classifier since we can simply apply a detector based on the 40\% negativity threshold and achieve 0.52 accuracy for cyberbullying detection.

\iffalse
For assigning one label to each data points as experiencing cyberbullying or not, we considered the confidence level given by CrowdFlower which is  a weighted version of majority voting method by "degree of trust". Degree of trust is calculated based on the ratio of the gold questions which has been answered correctly by the contributor. If an instance has confidence level greater than or equal to 60\%, we will consider that as strong enough sample of cyberbullying. In fact here we are running the classifier for strongest case, but we are not rejecting labeled samples with less 60\% as not having cyberbullying. With this criteria, 90\% of the labels obtained from second question meet the required confidence. In this data set, 45\% of the instances belong to class "not-bullying" and 55\% belongs to "bullying" group.
\fi

Two types of features were evaluated, namely those features obtained from the content of comments, and those features obtained from shared media objects and the profile owner.  For the text features, first we applied a pre-processing step to remove characters such as ``!", ``>'', etc. and stop words such as ``and",``or",``for", etc.   Features extracted from text include unigram, bigram, 3-gram,\iffalse POS (part of speech tag),\fi number of comments for the image, and number of posts within interval less than one hour.  Features extracted from user and media information (named as meta data) includes the number of followed-by's, follows, likes, and shared medias and features extracted from image content includes image categories.

\iffalse
Features extracted from text are:
\begin{itemize}
 \item unigram
  \item 3-gram
 \item POS (Part of Speech Tag)
  \item Number of comments for the image
  \item Number of posts with interval less than 1 hour
  \item Image category
  \item Followed by
\end{itemize}

Features extracted from text are:
\begin{itemize}
 \item Unigram
 \item Bigram
  \item 3-gram
 \item POS (Part of Speech Tag)
 \item Number of capital words
\end{itemize}

Features extracted from user information, media information and image content are:
\begin{itemize}
  \item Number of comments for the image
  \item Number of likes for the image
  \item Number of posts with interval less than 1 hour
  \item Image categories
  \item Followed by
   \item Follows
   \item Posting intervals
\end{itemize}

For our classifier which Na\"{i}ve Bayes Classifier as a basic simple and fast classifier suitable that works well for small datasets. 
\fi

Table~\ref{table:class1} illustrates the best performance results among different examined classifiers (all numbers are average over 10-fold cross validation results).  In the first row  using low dimensional feature space of meta data and a simple  Na\"{i}ve Bayes Classifier we jumped to accuracy 0.71 from baseline 0.52. Next we observed that adding image categories increased the accuracy to 0.72, with a high recall 0.78.  

In another experiment, only the text features unigram and 3-gram gave us the best accuracy using linear Support Vector Machine (SVM) Classifier. However, the dimension of unigrams and 3-gram features is very high, so next row shows the accuracy after applying Singular Value Decomposition (SVD) on text features. We observed keeping only the first 200 components, we can get the same accuracy. 

In the next step we added meta data and image categories to the text features. To get the best accuracy, we first standardized these set of features, applied kernel PCA (Principle Component Analysis) and kept the first 20 components. Then we concatenate this set of reduced dimension features with the reduced dimension features obtained from text. Applying linear SVM classifier, the accuracy jumped to 0.87 with   both high precision and recall. 

\emph{In summary, by employing multi-modal features obtained from text, meta data and images as input into a linear SVM classifier,  the accuracy of cyberbullying detection was meaningfully improved by 0.35 to a total of 0.87 compared to a base case of 0.52.  Simple meta data features gain accuracy 0.71, but to increase recall, more complex features are needed.}

\begin{table*}[htbp!]
\centering
\begin{tabular}{ |l | c | c | c | c |  }
\hline 
    \hline
      Features & Classifier & Accuracy & Precision & Recall     \\ \hline
      Meta data  & Na\"{i}ve Bayes & 0.71 &  0.75 & 0.66 \\ \hline
     Meta data, image categories & Na\"{i}ve Bayes & 0.74 &  0.74 & 0.78 \\ \hline
   Unigram, 3-gram &  linearSVM  & 0.85 &  0.88 & 0.84 \\ \hline
  SVD + Unigram, 3-gram &   linearSVM    & 0.85 &  0.84 & 0.88 \\ \hline
   SVD  +(Unigram, 3-gram), kernelPCA+(meta data, image categories)&     linearSVM & 0.87 &  0.88 & 0.87 \\ \hline
    \hline
\end{tabular}
\caption{Cyberbullying detection's classifier performance}
\label{table:class1}
\end{table*}

% !TEX root = main.tex
\section{Discussion and Future Work}

One theme for future work is to improve the performance of our classifier by adding more input features, such as new image features, temporal behavior of commenting, mobile sensor data, etc.  A limitation of our current classifier is that it is designed only for highly negative media sessions.  A more general classifier that can apply to all media sessions is needed.  This will also require us to enlarge our labeled data set substantially.  Incorporating image features needs to be automated by applying image recognition algorithms.  We plan to explore this research direction as well.  We have applied a majority vote definition in designing our classifier.  Another definition to consider is when at least one labeler has declared that he/she thinks this media session constitutes cyberbullying.  New classifiers will have to be designed for this definition.  

We also plan to consider designing classifiers for cyberaggression in addition to cyberbullying, and to investigate those media sessions that represent the former but not the latter behavior.

Another theme for future work is to obtain greater detail from the labeling surveys.  Our experience was that streamlining the survey improved the response rate, quality and speed.  However, we desire more detailed labeling, such as for different roles in cyberbullying -- identifying and differentiating the role of a victim's defender, who may also spew negativity, from a victim's bully or bullies.

\iffalse
Improve classifier accuracy with other features.  We are using most negative, may need to randomly sample more normal users.  Different classifier algorithms.

Designing the right survey is hard.  More detailed survey.

Cyberaggression classifier - gray area - only not cyberbullying

Cyberbullying extending definition to less than 3 with cyberaggression support strong, it is cyberbullying if at least one, ...

Are there automated image labeling...

How objective is cyberaggression vs cyberbullying, subjective, ...

automated detection?  automated extraction of tattoo, ... use existing image recognition, ... other features readily available, ... or use other automated image features, 

potentially a lot more features we can extract from the images, ...
\fi
% !TEX root = main.tex
\section{Conclusions}

We believe this paper makes the following major contributions: an appropriate definition of cyberbullying that incorporates both frequency of negativity and imbalance power is applied in large-scale labeling, and is differentiated from cyberaggression; cyberbullying is studied in the context of a media-based social network, incorporating both images and comments in the labeling; a detailed analysis of the distribution results of the labeling of cyberbullying incidents is presented, including a correlation analysis of cyberbullying with other factors derived from images, text comments, and social network meta data; multi-modal classification results are presented that incorporate a variety of features to identify cyberbullying incidents.

The major findings of this paper comprise the following results.  First, a key finding of our labeling is that about 48\% of Instagram media sessions were not deemed as cyberbullying using a majority vote criterion among five labelers, even though these were among the media sessions with the highest percentage of profanity words, i.e. a significant fraction of negative content does not constitute acts of online cyberbullying.  Second, labelers are mostly in agreement about what behavior constitutes cyberbullying and what does not in Instagram media sessions.  Third, our analysis identified that that there is significant class of Instagram media sessions that exhibits cyberaggression but not cyberbullying.  Fourth, there are strong correlations between the strength of support for labeled cyberbullying and the number of text comments as well as the temporal
property of the number of comments that are posted within one hour of one another in an Instagram media session.  Fifth, we demonstrate that a Linear SVM classifier can significantly improve the accuracy of identifying cyberbullying to 87\% by incorporating multi-modal features from text, images, and meta data for the media session.

\iffalse
What are the major findings?

1) about 1/3 of most negative images+comments were *not* cyberbullying, so a pure list-based approach is not sufficiently accurate \\
2) able to achieve x\% accuracy \\
3) which features make the most difference? \\
4) add more here...

1.	Not all negative content constitutes acts of online cyberbullying. In fact, about 1/3 of the most negative images and comments were not characterized as cyberbullying, therefore a pure list-based approach of negative words are not sufficient for detecting cyberaggression and cyberbullying.
2.	Cyberbullying detection in OSN’s is possible with an X\% accuracy. These results provide promise for creating a system that will auto-detect acts of online cyberaggression and cyberbullying and a method for intervening with the victims. 
3.	Images of people and text are most likely to elicit cyberbullying comments. XXis this correct?XX This finding will help design the classifiers needed to efficiently and effectively identify cyberaggression and cyberbullying in OSN’s.
\fi

\bibliographystyle{aaai}
\bibliography{homa}
\iffalse
\section{ Acknowledgments}
 We wish to acknowledge financial support for this research from the US National Science Foundation (NSF) through grant CNS 1162614.
\fi

\end{document}